\documentclass[conference]{IEEEtran}
\IEEEoverridecommandlockouts
\usepackage{cite}
\usepackage{amsmath,amssymb,amsfonts}
\usepackage{algorithmic}
\usepackage{graphicx}
\usepackage{textcomp}
\usepackage{xcolor}
\usepackage{url}
\usepackage{hyperref}
\def\BibTeX{{\rm B\kern-.05em{\sc i\kern-.025em b}\kern-.08em
    T\kern-.1667em\lower.7ex\hbox{E}\kern-.125emX}}
\begin{document}

\title{PLUTUS OPEN SOURCE\\Breaking Barriers in Algorithmic Trading}

\author{\IEEEauthorblockN{An-Dan Nguyen}
\IEEEauthorblockA{\textit{ALGOTRADE}\\
Ho Chi Minh City, Vietnam \\
andan@algotrade.vn}
\and
\IEEEauthorblockN{Quang-Khoi Ta}
\IEEEauthorblockA{\textit{ALGOTRADE}\\
Ho Chi Minh City, Vietnam \\
khoi.ta@algotrade.vn}
\and
\IEEEauthorblockN{Duy-Anh Vo}
\IEEEauthorblockA{\textit{ALGOTRADE}\\
Ho Chi Minh City, Vietnam \\
anh.vo@algotrade.vn}}

\maketitle

\begin{abstract}
Algorithmic trading has long been an opaque, fragmented domain, guarded by secrecy and built around proprietary systems. In contrast to the open, collaborative evolution in fields like machine learning or software engineering, the algorithmic trading ecosystem has been slow to adopt reproducibility, standardization, and shared infrastructure.

This paper introduces PLUTUS Open Source, an initiative sponsored by ALGOTRADE to reshape this landscape through openness, structure, and collaboration. PLUTUS combines a reproducibility standard, a modular development framework, and a growing suite of community-built reference strategies. The project provides a systematic approach to designing, testing, and documenting trading algorithms, regardless of the user's technical or financial background.

We outline the motivation behind the initiative, present its foundational structure, and showcase working examples that adhere to the PLUTUS standard. We also invite the broader research and trading communities to contribute, iterate, and help build a transparent and inclusive future for algorithmic trading.
\end{abstract}


\section{Introduction}
Closed doors, secret signals, and proprietary systems have long defined algorithmic trading. While other fields, such as artificial intelligence and data science, have benefited from open-source collaboration, reproducible research, and community-driven innovation, algorithmic trading remains an outlier. Fragmented, opaque, and often hostile to newcomers, the field has mainly progressed in isolation, reinforcing a cycle of exclusivity and secrecy.

PLUTUS Open Source is an effort to change that reality.

Sponsored by ALGOTRADE, PLUTUS is not just another open-source repository. It is a structured, community-first initiative to bring standardization, reproducibility, and accessibility to a field that has traditionally resisted all three. The goal is to reimagine what algorithmic trading could look like if the barriers to entry were lowered, best practices were codified, and high-quality research became both replicable and extensible.

PLUTUS introduces a reproducibility standard, a modular development framework, and a living repository of sample strategies, each fully documented, version-controlled, and aligned with a 9-step development process. But more importantly, PLUTUS is a platform for learning and contribution: a bridge for programmers without finance backgrounds and finance professionals without engineering skills.

This paper introduces the motivations and structure behind PLUTUS Open Source. We explore its reproducibility framework, present working examples, and outline how contributors can engage with and expand the initiative. If you've ever wondered how to build credibility in a field built on secrecy, or how to start contributing meaningfully to its future, this paper is your blueprint.

\section{Motivation}
Despite algorithmic trading's technological sophistication, the field remains largely inaccessible to those outside tightly controlled professional circles. The lack of openness is not just cultural; it's structural. There is no shared framework for development, no common language for collaboration, and no unified pathway for newcomers to learn and contribute meaningfully.

At the same time, the demand for such a structure has never been greater. The rise of quantitative finance, the availability of data, and the global reach of digital markets have drawn a new generation of builders to the space: developers, data scientists, mathematicians, engineers, and finance professionals. Each brings a unique perspective, but also faces unique barriers.

We observe two dominant audience types entering the field:
\begin{itemize}
    \item Tech learners: Individuals with strong programming and analytical backgrounds, often proficient in machine learning, statistics, or systems design. These users can automate workflows and scale infrastructure, but often lack financial domain knowledge, making it challenging to form sound trading hypotheses or interpret market behavior.
    \item Finance learners: Investors, traders, students, and professionals trained in market dynamics, behavioral finance, or investment strategy. These users can form promising ideas, but often lack the engineering skills to test, scale, or validate their approaches algorithmically.
\end{itemize}

PLUTUS Open Source is designed to bridge this divide.
\begin{itemize}
    \item For technical users, PLUTUS provides a clear structure, reproducibility standards, and tooling to build and iterate on strategies quickly, accelerating their learning curve and surfacing real signals from noise. 
    \item For finance-native users, it offers ready-made automation infrastructure and reproducible templates, reducing the friction involved in transforming investment logic into working code.    
\end{itemize}

However, the motivation for PLUTUS goes beyond onboarding. It is about credibility. Reproducibility becomes the cornerstone of trust in a field where performance often hides behind cherry-picked backtests and unverifiable claims. A standardized, community-maintained framework ensures that strategies can be shared, discussed, stress-tested, and improved like code in any mature open-source discipline.

PLUTUS is not trying to replace proprietary edge. It is trying to raise the floor.

The more projects that follow shared processes, disclose testing details, and invite peer scrutiny, the faster the field matures, and the more inclusive, credible, and innovative it becomes.

\section{Background and Related Works}
The landscape of algorithmic trading research has historically operated in silos, split between proprietary institutional models and fragmented academic output. Industry systems tend to be closed, inaccessible, and tied to confidential infrastructure. While often innovative, academic studies typically suffer from reproducibility limitations due to unavailable datasets, undocumented assumptions, or missing source code \cite{bailey2014overfitting}\cite{ioannidis2005false}.

This is not unique to finance. Across the broader scientific community, a growing body of literature has documented what is now widely referred to as the reproducibility crisis, the realization that many published findings cannot be independently verified, even by experts within the same field \cite{openscience2015psych}\cite{baker2016repro}. Factors contributing to this problem include poor documentation, selection bias, data dredging, and the lack of incentives for publishing negative or null results.

By contrast, machine learning and data science fields have rapidly evolved through open-source ecosystems. Platforms like TensorFlow\cite{abadi2016tensorflow}, PyTorch\cite{paszke2019pytorch}, and scikit-learn\cite{pedregosa2011scikit} have demonstrated how community-driven frameworks, standardized pipelines, and shared benchmarks accelerate progress and lower entry barriers. These projects have also formalized reproducibility norms, enabling collaborative validation and iteration across institutions.

Several efforts have attempted to bring similar openness to quantitative trading:
\begin{itemize}
    \item QuantConnect\cite{quantconnect2024} and Zipline\cite{zipline} provide robust infrastructure for backtesting and execution, but their focus is global and generic. They do not enforce reproducibility standards or cater to emerging markets with unique structural constraints.
    \item Backtrader\cite{backtrader2024}, bt, and other Python libraries offer basic simulation and portfolio modeling functionality, but lack standardization in documentation, results reporting, or research transparency.
    \item Many academic papers introduce promising strategy formulations without accessible datasets or executable code, limiting their practical value and undermining replicability\cite{lopez2018advances}.
\end{itemize}
More importantly, none of these tools explicitly address the unique needs of localized markets, such as Vietnam's stock exchange, where data availability, liquidity structure, and institutional behaviors differ significantly from U.S. or European contexts. As a result, researchers and retail participants in these markets are left to reverse-engineer infrastructure or rely on closed platforms that offer little transparency or extensibility.

What's missing is a framework that acts more like a protocol than a toolkit, one that enforces structure while remaining flexible, supports localized market conditions, and bridges the gap between raw experimentation and verifiable results.

PLUTUS Open Source is designed to fill that void.

Built from the ground up with reproducibility, modularity, and accessibility as core design principles, PLUTUS provides not only infrastructure but a shared language for strategy research. It draws inspiration from DevOps practices and open science methodology, applying them to algorithmic trading with a focus on regional inclusion and cross-disciplinary participation.

\section{The PLUTUS Initiative}
The PLUTUS Open Source\footnote{\href{https://github.com/algotrade-plutus}{https://github.com/algotrade-plutus}} project was launched with a clear, practical vision: to bring openness, reproducibility, and standardization to a field traditionally dominated by secrecy and fragmentation. Sponsored by ALGOTRADE, PLUTUS is not just an infrastructure tool but a cultural initiative aimed at transforming how algorithmic trading research is created, shared, and evaluated.

At its heart, PLUTUS is designed to answer one core question:

\textit{What would algorithmic trading look like if it embraced the open-source principles that transformed fields like machine learning and software development?}

To answer that, PLUTUS is built around four core objectives:

\subsection{The Four Core Objectives of PLUTUS}
\subsubsection{Standardize the Practice}
PLUTUS provides a unified framework of tools, design patterns, and naming conventions so developers, researchers, and traders can speak the same language. Making high-quality practices repeatable reduces the learning curve for newcomers and increases interoperability across teams.

\subsubsection{Lower the Entry Barrier}
Through structured templates, documentation, and reference implementations, PLUTUS offers a guided on-ramp for people entering the field, whether from finance or tech. The framework enables individuals to go from concept to prototype without reinventing infrastructure.

\subsubsection{Enable Reproducibility and Transparency}
PLUTUS introduces clear expectations for project documentation and verification. Each project can be tested and audited using a consistent methodology, reducing ambiguity and reinforcing the credibility of results.

\subsubsection{Promote Innovation and Fairness}
By open-sourcing high-quality strategies, workflows, and educational materials, PLUTUS helps level the playing field. It invites participation from students, independent quants, and researchers around the world, accelerating experimentation and idea exchange across borders and backgrounds.

\subsection{The PLUTUS Standard}
The PLUTUS Standard is a set of conventions, documentation requirements, and evaluation metrics designed to ensure that trading research is understandable, testable, and reproducible. It draws from practices standard in open-source software engineering, such as templated README structures, version control discipline, and result reproducibility. It adapts them to the workflow of algorithmic strategy development.

Central to this standard is the PLUTUS Compliance Score, a rubric that grades a project's adherence to reproducibility best practices. Projects are encouraged to present results using a consistent structure, including documented assumptions, configuration files, code execution steps, and performance reporting.

\subsection{The Framework (and Future Platform)}
Beyond guidelines, PLUTUS offers an extensible framework to accelerate the research and development lifecycle. This includes:
\begin{itemize}
    \item Modular pipelines for each phase of the 9-step development process
    \item Strategy templates that are plug-and-play for backtesting, optimization, and validation
    \item Integration support for paper trading, performance analytics, and deployment tools
\end{itemize}
The framework acts as the glue between ideas and implementation, giving both individuals and teams a structured, extensible environment to build within. Over time, this foundation will evolve into a full-featured platform that supports continuous strategy iteration, collaboration, and deployment.

\subsection{The Process and Knowledge Base}
Finally, PLUTUS is a process. It promotes not just tools, but a shared way of thinking about research. The initiative provides:
\begin{itemize}
    \item Educational materials for practitioners and students
    \item Contributor guides and project templates for new teams
    \item A growing repository of PROTO sample projects that follow the standard
\end{itemize}

This knowledge base is continuously expanded by the community, ensuring that PLUTUS remains accessible, current, and useful at every level of experience, from first-time traders to institutional quants.

\subsection{Who PLUTUS Is For}
PLUTUS is designed to meet users where they are. It's not exclusive to elite quants or professional developers. It is built for:
\begin{itemize}
    \item Technologists (programmers, data scientists, statisticians) who want to explore trading but lack domain experience
    \item Finance professionals (traders, investors, fund managers, students) who want to codify and test strategies at scale
    \item Educators and researchers looking for a transparent way to publish, validate, and extend their findings
\end{itemize}

Whether building your first bot or formalizing a multi-strategy portfolio, PLUTUS offers a structure that scales with your ambition.

\section{The Reproducibility Standard of PLUTUS}
Reproducibility is foundational to credible scientific progress, but it is often lacking in algorithmic trading. Strategy write-ups rarely include code, backtests are frequently non-verifiable, and even well-meaning academic research usually lacks the documentation or environment controls needed to replicate results.

PLUTUS addresses this head-on by introducing a Reproducibility Standard tailored to the realities of trading system development. It sets clear expectations for documentation, execution paths, and result validation. The goal is to make independent verification not only possible but routine.

\subsection{Scope of the Standard}
The PLUTUS standard\footnote{\href{https://github.com/algotrade-plutus/plutus-guideline}{https://github.com/algotrade-plutus/plutus-guideline}} focuses on reproducibility through Step 7: Paper Trading of the 9-Step Development Process for Trading Strategies \footnote{\href{https://hub.algotrade.vn/knowledge-hub/steps-to-develop-a-trading-algorithm/}{https://hub.algotrade.vn/knowledge-hub/steps-to-develop-a-trading-algorithm/}}. A project is considered reproducible under PLUTUS if:
\begin{itemize}
    \item All results can be replicated using the provided source code and instructions
    \item No direct communication with the author is needed to execute or validate the system
    \item Reported results in the documentation match the outputs when the code is run independently
    \item Data sources, processing steps, and parameter settings are fully disclosed
\end{itemize}

This standard is enforced through a structured repository format, a reproducibility checklist, and an optional scoring system to evaluate compliance.

\subsection{Required Project Structure}
Every PLUTUS-compliant repository must contain a $README.md$ file that follows a standard structure and provides a full blueprint for reproducing the project's findings. The required sections include:
\begin{itemize}
    \item Abstract: A summary of the hypothesis, methods, and key results
    \item Introduction: Motivation, goals, and overview
    \item Related Work (optional): Prior strategies or concepts informing the project
    \item Trading Hypotheses (Step 1): Description of the edge being pursued
    \item Data (Steps 2–3): Collection, processing, and structure
    \item Implementation (Steps 4–6+): How to run, configure, and test the code
    \item Backtesting \& Optimization: In-sample and out-of-sample result reporting
    \item Paper Trading (optional Step 7): Results from simulated real-time deployment
    \item Conclusion (optional): Final observations or future ideas
    \item References: Datasets, APIs, libraries, or academic sources
    \item Final Report or Paper (optional): Deeper methodology and results discussion
\end{itemize}

Each section is not just a placeholder; it is meant to provide concrete, step-by-step instructions for full project replication.

\subsection{Why This Matters}
Every repo is its own little universe without a standard, with different structures, inconsistent terminology, and unclear replication paths. PLUTUS fixes that. The goal is not to constrain creativity but to standardize the fundamentals, just like version control, unit testing, and style guides do in software.

Reproducibility builds credibility. It's the foundation for discussion, collaboration, and innovation. The more projects that follow the PLUTUS Standard, the richer and more useful the entire ecosystem becomes.

We do not just want working strategies; we want strategies you can verify, tweak, extend, and build on. That's how fields move forward.

And in trading, moving forward is everything.

\subsection{Bridging into Practice}
The next two sections present real-world projects developed under the PLUTUS standard to illustrate reproducibility in action. These are not toy examples. They are fully replicable strategies tested across multiple phases, documented in detail, and shared openly for reuse and adaptation.

Together, they form the foundation of the PROTO series: a growing collection of reference implementations designed to serve as templates for real-world, production-ready trading research.

\section{PROTO: Smart Beta}
This section presents a working reference implementation: PROTO: Smart Beta\footnote{\href{https://github.com/algotrade-plutus/ProtoSmartBeta}{https://github.com/algotrade-plutus/ProtoSmartBeta}} to bring the PLUTUS standard to life. This strategy exemplifies how reproducibility, structure, and transparency can be applied to a long-term, factor-based investing strategy, making it functional, shareable, testable, and extensible.

PROTO: Smart Beta was developed using the full PLUTUS development process and published with complete documentation and code under the standard repository structure. Its goal is not to outperform every benchmark, but to serve as a clean, minimal template for building and validating systematic investing strategies.

\subsection{Trading Hypothesis}
This strategy is built on classic value investing principles: firms with lower valuation ratios and consistent dividend payouts are more likely to outperform over the long term.
Two signals are used to screen the equity universe:
\begin{itemize}
    \item Low Price-to-Earnings (P/E) Ratio
    \item Positive Dividend Yield (DY)
\end{itemize}
The core hypothesis is that by filtering for undervalued and income-generating stocks, one can construct a stable, outperforming portfolio in inefficient or emerging markets.

\subsection{Strategy Rules}
At the end of each month:
\begin{itemize}
    \item All current positions are exited.
    \item Stocks are screened based on the following:
    \begin{itemize}
        \item P/E ratio between 0 and 15
        \item DY greater than 0.01
    \end{itemize}
    \item The portfolio is rebalanced equally among all qualified stocks.
    \item A transaction fee of 0.035\% is applied on each buy and sell.
    \item Risk-free rate is set at 6\% annually
    \item The chosen benchmark is VNINDEX
\end{itemize}

The rules are intentionally minimal to emphasize transparency and ease of replication. All logic is implemented via parameterized functions and configuration files.

\subsection{Data}
\begin{itemize}
    \item Source: Algotrade internal database
    \item Period: Jan 1, 2022 – Jan 1, 2025
    \item Asset Universe: Vietnamese listed equities
    \item Tools: SQL-based data collection, Python preprocessing
\end{itemize}

Raw data is filtered for missing values, aligned monthly, and exported as .csv files for backtesting pipelines. Feature columns include financial statement metrics and market prices.

\subsection{In-Sample Backtesting}
\begin{itemize}
    \item Period: Jan 1, 2019 – Jan 1, 2022
    \item Benchmark: VNINDEX
    \item Frequency: Monthly rebalancing
\end{itemize}

\subsubsection{Performance Metrics}
\begin{table}[ht]
\centering
\begin{tabular}{ll}
\textbf{Metric}        & \textbf{Value} \\
Sharpe Ratio           & 1.2971         \\
Information Ratio      & 0.0298         \\
Sortino Ratio          & 1.7297        \\
Maximum Drawdown (MDD) & -28.28\%      
\end{tabular}
\caption{In-sample Backtesting Performance}
\label{tab:smart-beta-in-sample}
\end{table}

These metrics show strong performance during the training period. The positive Sharpe and Sortino ratios suggest favorable risk-adjusted returns, with adequate downside protection. Although the information ratio remains modest, it indicates that the strategy has some predictive power over the benchmark. The drawdown, while notable, is within acceptable bounds for a monthly-rebalanced equity strategy in an emerging market.

\subsubsection{Charts}
The following charts provide a visual overview of the strategy's in-sample performance. The NAV curve shows the equity trajectory of the strategy, while the drawdown curve illustrates peak-to-trough losses during the period.
\begin{figure}[ht]
    \centering
    \includegraphics[width=\columnwidth]{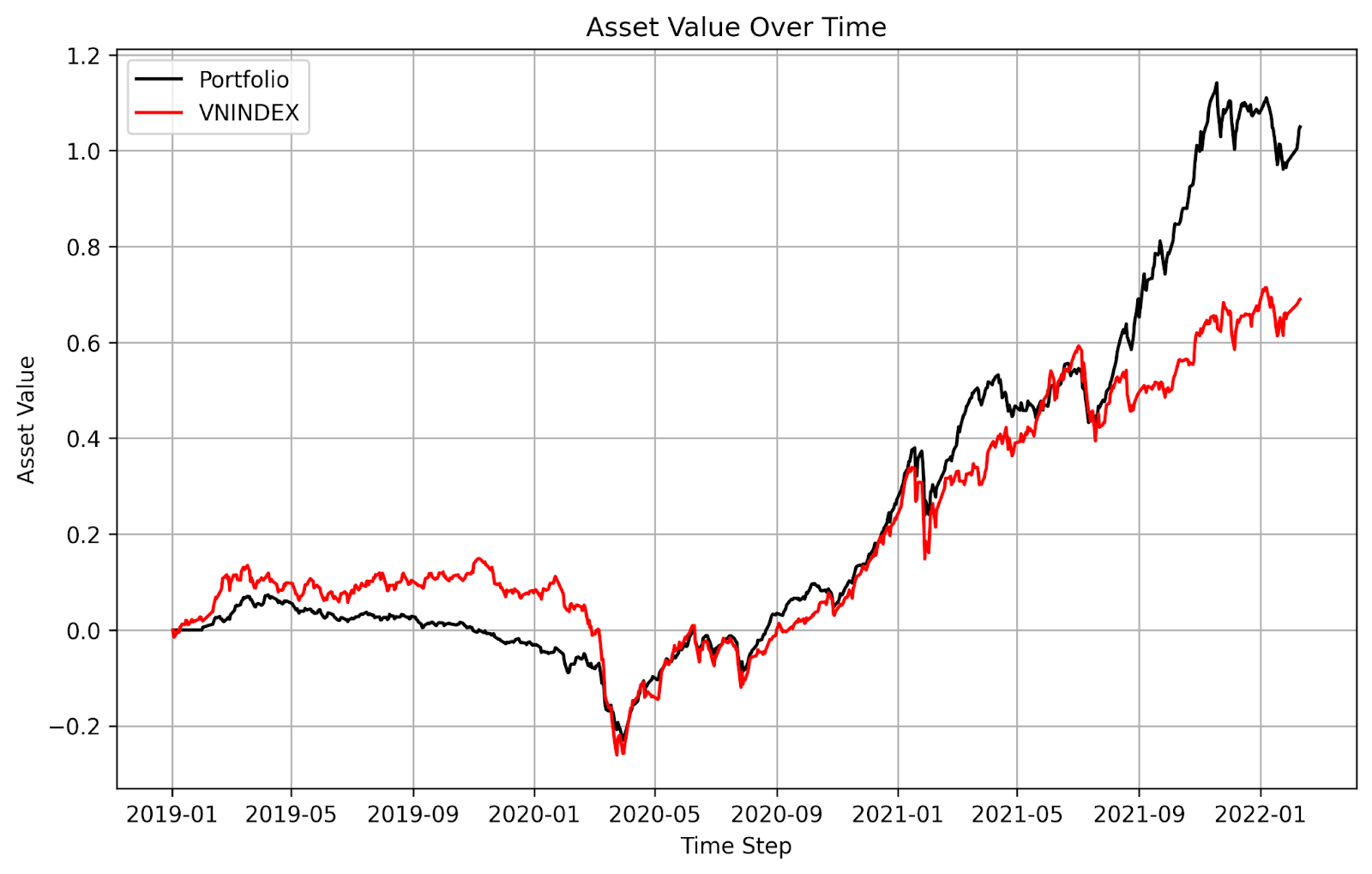}
    \caption{Smart Beta NAV Chart In-sample}
    \label{fig:nav-in-sm}
\end{figure}
\begin{figure}[ht]
    \centering
    \includegraphics[width=\columnwidth]{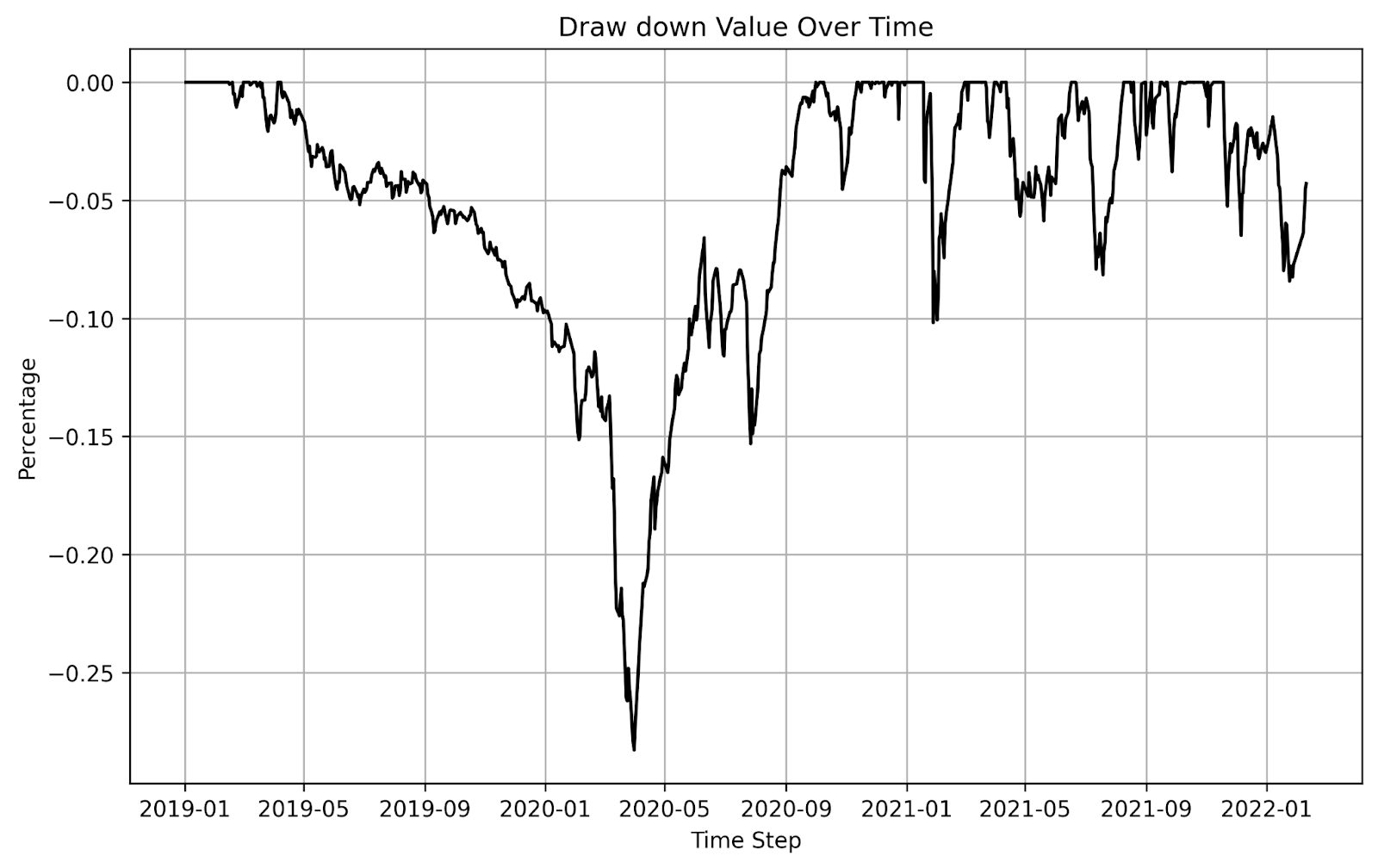}
    \caption{Smart Beta Drawdown Chart In-sample}
    \label{fig:dd-in-sm}
\end{figure}
Output was tested on multiple runs to ensure determinism.

\subsection{Optimization}
The optimization process focused on identifying the best range bounds for the P/E ratio and Dividend Yield to fine-tune the strategy's screening criteria. These ranges define the eligibility window for stocks at each rebalance.

The optimization was performed using Optuna, a modern hyperparameter optimization framework. The objective function was to maximize the Sharpe Ratio of the resulting portfolio over the in-sample training period, subject to diversification and data sufficiency constraints.

\begin{table}[!ht]
\centering
\begin{tabular}{lll}
\textbf{Parameter} & \textbf{Lower Bound} & \textbf{Upper Bound}\\
P/E Ratio              & 0  & 15 \\
Dividend Yield         & 0.01  & $+\infty$ \\
\end{tabular}
\caption{Smart Beta: Optimized Parameters}
\label{tab:smart-beta-optim-params}
\end{table}

A stock qualifies if:
\begin{itemize}
    \item Its P/E ratio is between 0 and 15
    \item Its dividend yield is greater than 1\%
\end{itemize}

These bounds reflect a traditional value investing logic: excluding speculative or overpriced stocks while filtering out firms without meaningful income payouts. The resulting configuration was used in the out-of-sample backtest.

\subsection{Out-of-Sample Backtesting}
\begin{itemize}
    \item Period: Jan 1, 2022 – Jan 1, 2024
    \item Configuration: Optimized thresholds from prior step
\end{itemize}

\subsubsection{Performance Metrics}
\begin{table}[ht]
\centering
\begin{tabular}{ll}
\textbf{Metric}        & \textbf{Value} \\
Sharpe Ratio           & -0.6420         \\
Information Ratio      & 0.0206         \\
Sortino Ratio          & -0.7986        \\
Maximum Drawdown (MDD) & -47.15\%     
\end{tabular}
\caption{Out-of-sample Backtesting Performance}
\label{tab:in-sample}
\end{table}
The out-of-sample results show a marked decline in performance, with negative Sharpe and Sortino ratios indicating poor risk-adjusted returns and asymmetric downside. However, the strategy remained reproducible and structurally sound across reruns. The Information Ratio, though weak, suggests limited relative signal persistence, meriting future refinements or market condition adjustments.

Although performance declines compared to the in-sample performance, the results show the strength of the chosen Smart Beta strategy when outperforming the benchmark in the out-of-sample periods. That reflects the strategy's nature. The sharp decrease in Sharpe and Sortino ratios can be rationalized through the benchmark's downtrend in the same period.

\subsubsection{Charts}
Below are the equity curve and drawdown trajectory for the out-of-sample testing window. These charts allow readers to observe changes in volatility, drawdown behavior, and recovery patterns post-optimization.
\begin{figure}[ht]
    \centering
    \includegraphics[width=\columnwidth]{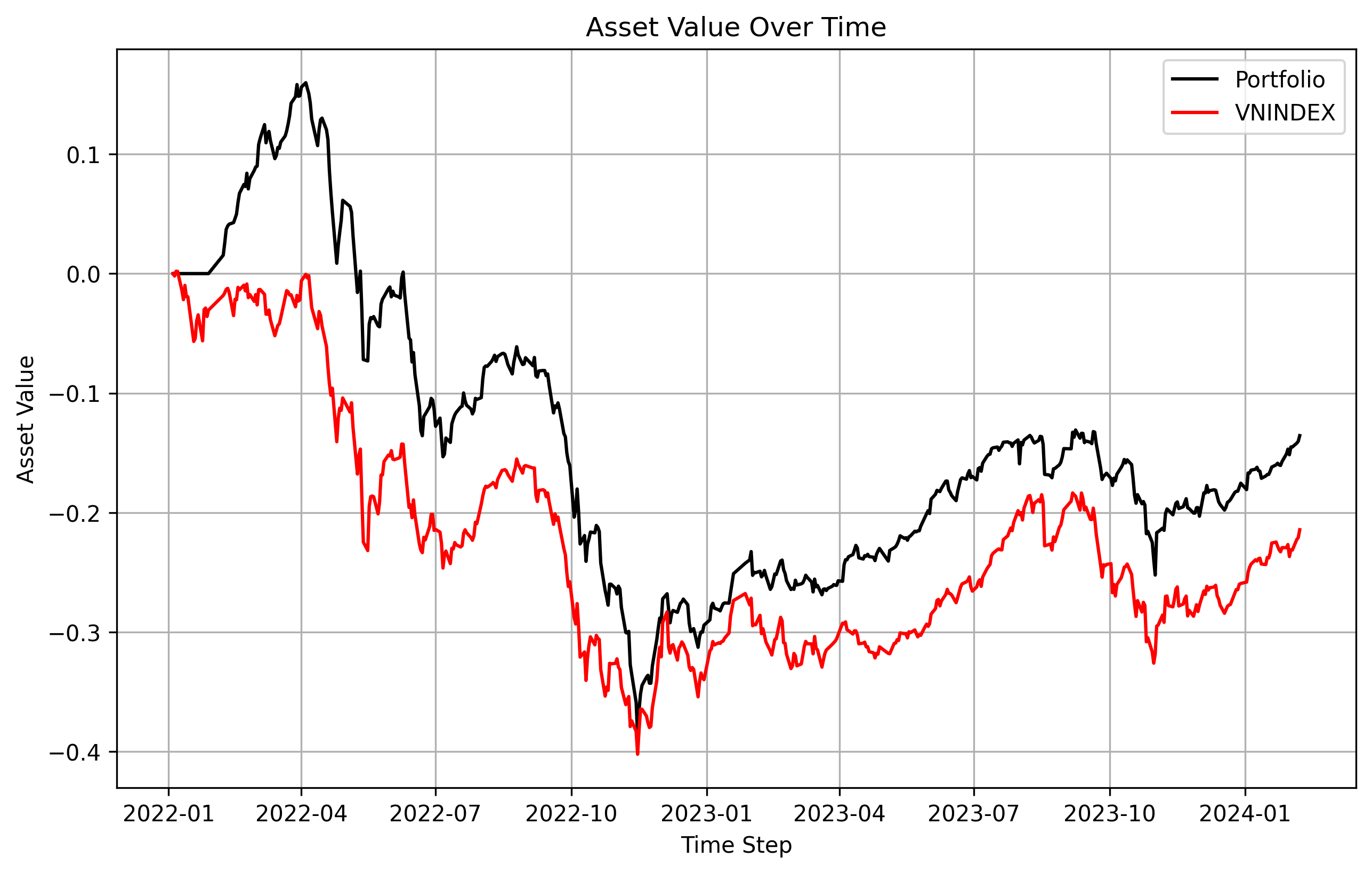}
    \caption{Smart Beta NAV Chart Out-of-Sample}
    \label{fig:nav-out-sm}
\end{figure}
\begin{figure}[ht]
    \centering
    \includegraphics[width=\columnwidth]{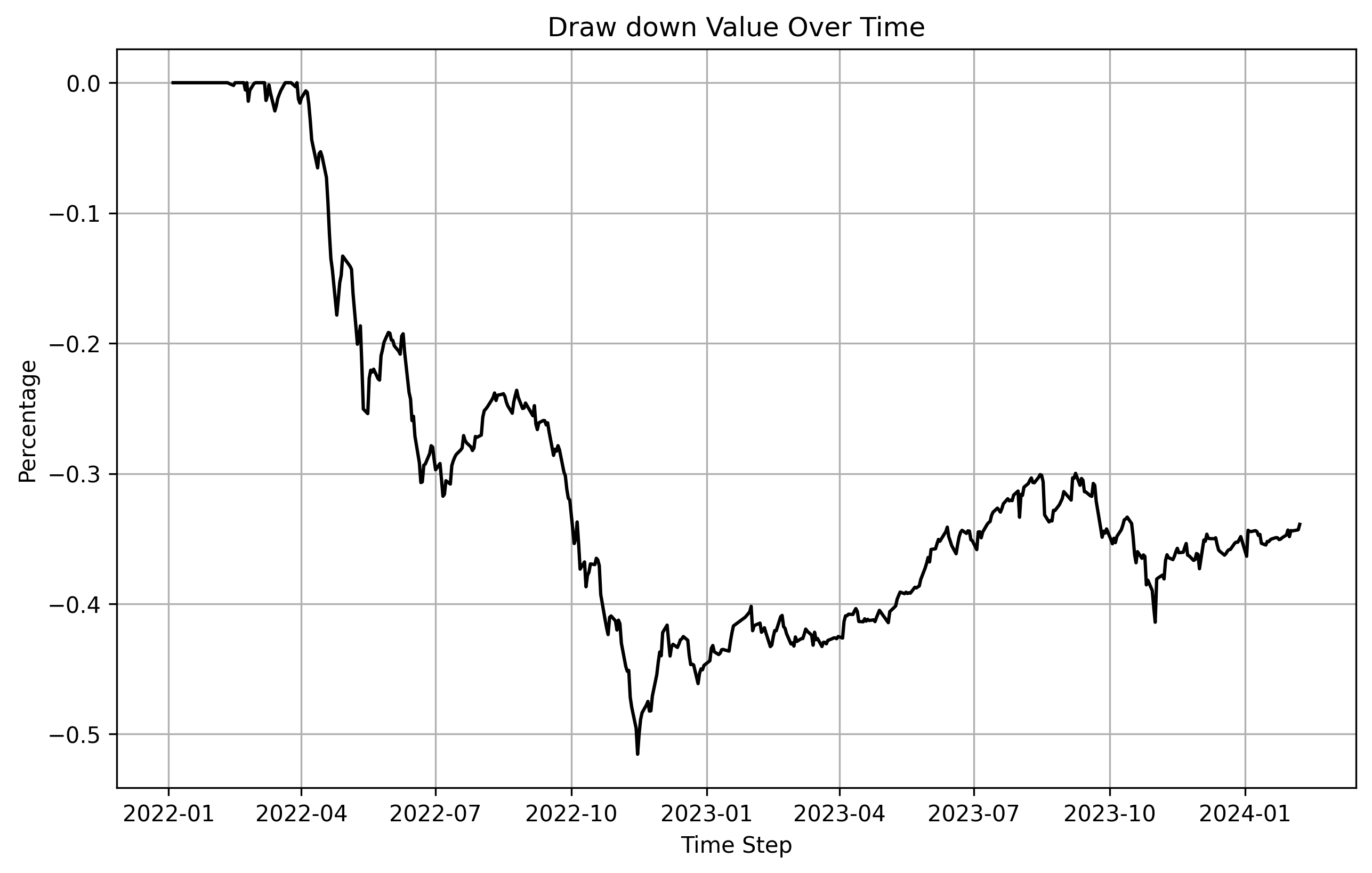}
    \caption{Smart Beta Drawdown Chart Out-of-Sample}
    \label{fig:dd-out-sm}
\end{figure}

\subsection{Paper Trading}
No live or delayed data paper trading was conducted for this implementation. All experiments were contained within historical simulations.

\subsection{Conclusion}
PROTO: Smart Beta shows how even traditional strategies can be elevated through structure and transparency. By focusing on simple signals, modular code, and thorough documentation, the strategy becomes a tool for learning, not just a performance product.

As part of the growing PROTO series, it provides a foundational template for building and refining Smart Beta portfolios using the PLUTUS standard.

\section{PROTO: Market Maker}
While PROTO: Smart Beta demonstrates PLUTUS in the context of long-term portfolio strategies, PROTO: Market Maker\footnote{\href{https://github.com/algotrade-plutus/ProtoMarketMaker}{https://github.com/algotrade-plutus/ProtoMarketMaker}} shifts focus to the microstructure domain, where trading decisions occur in seconds, not months.

Market making is one of the most essential forms of algorithmic execution. It involves continuously posting buy and sell limit orders, capturing spread while managing inventory risk. PROTO: Market Maker exemplifies how even low-latency strategies can be built and shared using the PLUTUS standard.

\subsection{Hypothesis}
The strategy operates under the assumption that inventory-aware quoting, adjusting bid and ask levels dynamically based on exposure, enables greater stability and risk control over time. When inventory becomes imbalanced (long or short), the system alters quote distances to encourage mean reversion toward neutrality.

This mechanism discourages directional drift, improves turnover distribution, and promotes long-term resilience even in volatile conditions.

\subsection{Strategy Rules}
The bid and ask prices are determined using the following formulas:
\begin{itemize}
    \item $bid = mid_{price} - step \times (\max(inventory, 0) \times 0.02 + 1)$
    \item $ask = mid_{price} - step \times (\min(inventory, 0) \times 0.02 - 1)$
\end{itemize}

Where:
\begin{itemize}
    \item Inventory is the net position
    \item Step is a configurable distance from mid-price
    \item Quotes are refreshed every 15 seconds or immediately upon a fill
    \item Each trade incurs a fee of 0.2 points, deducted from the execution price
\end{itemize}

This design allows the strategy to rebalance exposure passively without relying on price prediction. Execution is rule-based, deterministic, and traceable.

There are some notes about the current implementation:
\begin{itemize}
    \item The strategy is currently demonstrated in the VN30F1M instrument. Hence, the fee is based on points (not percentage)
    \item The initial capital to test the strategy is 500M Vietnamese Dong.
\end{itemize}

\subsection{Data}
\begin{itemize}
    \item Source: Algotrade internal database
    \item Period: Jan 1, 2022 – Apr 29, 2025
    \item Structure: Tick-level close prices, bid-ask spreads, execution logs
    \item Cleaning: Gaps forward-filled; timestamp normalization applied
\end{itemize}
This dataset enables near real-time simulation with sufficient market depth granularity to evaluate inventory dynamics and quote placement efficiency.

\subsection{In-Sample Backtesting}
\begin{itemize}
    \item Period: Jan 1, 2022 – Jan 1, 2023
    \item Execution: Every 15 seconds
    \item Inventory: No daily reset
    \item Fee Model: 0.2 points per trade
    \item Step Size: 1.8
\end{itemize}

\subsubsection{Performance Metrics}
The in-sample results indicate strong risk-adjusted performance under stable market conditions. The high Sharpe and Sortino ratios reflect consistent, high-quality returns with minimal downside volatility. Given the high-frequency nature and inventory exposure, the drawdown of $\sim 19\%$ is acceptable.

\begin{table}[!ht]
\centering
\begin{tabular}{ll}
\textbf{Metric}        & \textbf{Value} \\
Sharpe Ratio           & 1.5619        \\
Sortino Ratio          & 2.3335        \\
Maximum Drawdown (MDD) & -18.91\%
\end{tabular}
\caption{In-sample Backtesting Performance}
\label{tab:market-maker-in-sample}
\end{table}

\subsubsection{Charts}
The NAV curve illustrates steady growth over the test period, while the drawdown curve shows moderate but recoverable losses. The inventory plot confirms that the quoting logic effectively managed exposure, frequently returning the strategy to near-neutral inventory.
\begin{figure}[ht]
    \centering
    \includegraphics[width=\columnwidth]{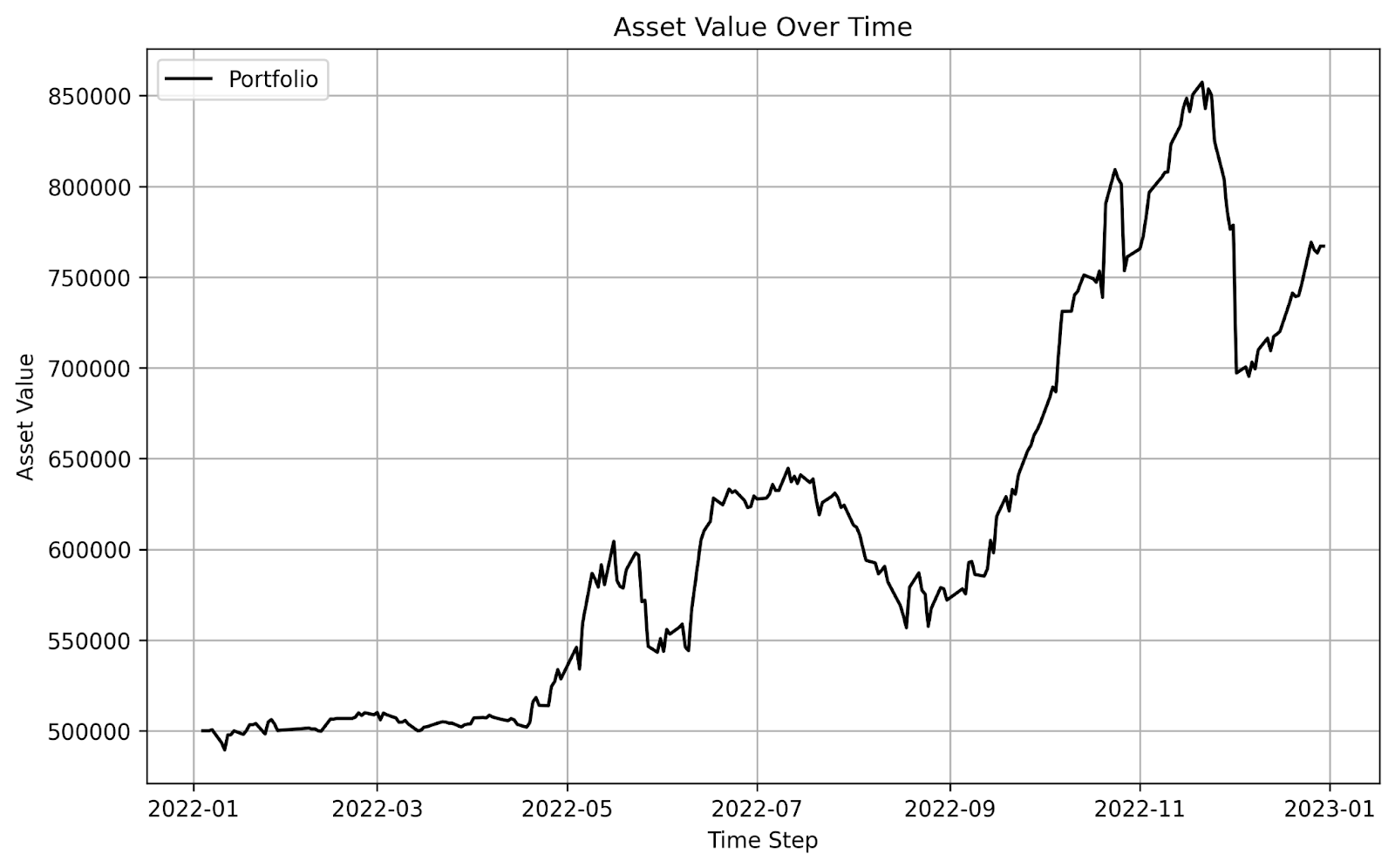}
    \caption{Market Maker NAV Chart In-sample}
    \label{fig:nav-in-mm}
\end{figure}
\begin{figure}[ht]
    \centering
    \includegraphics[width=\columnwidth]{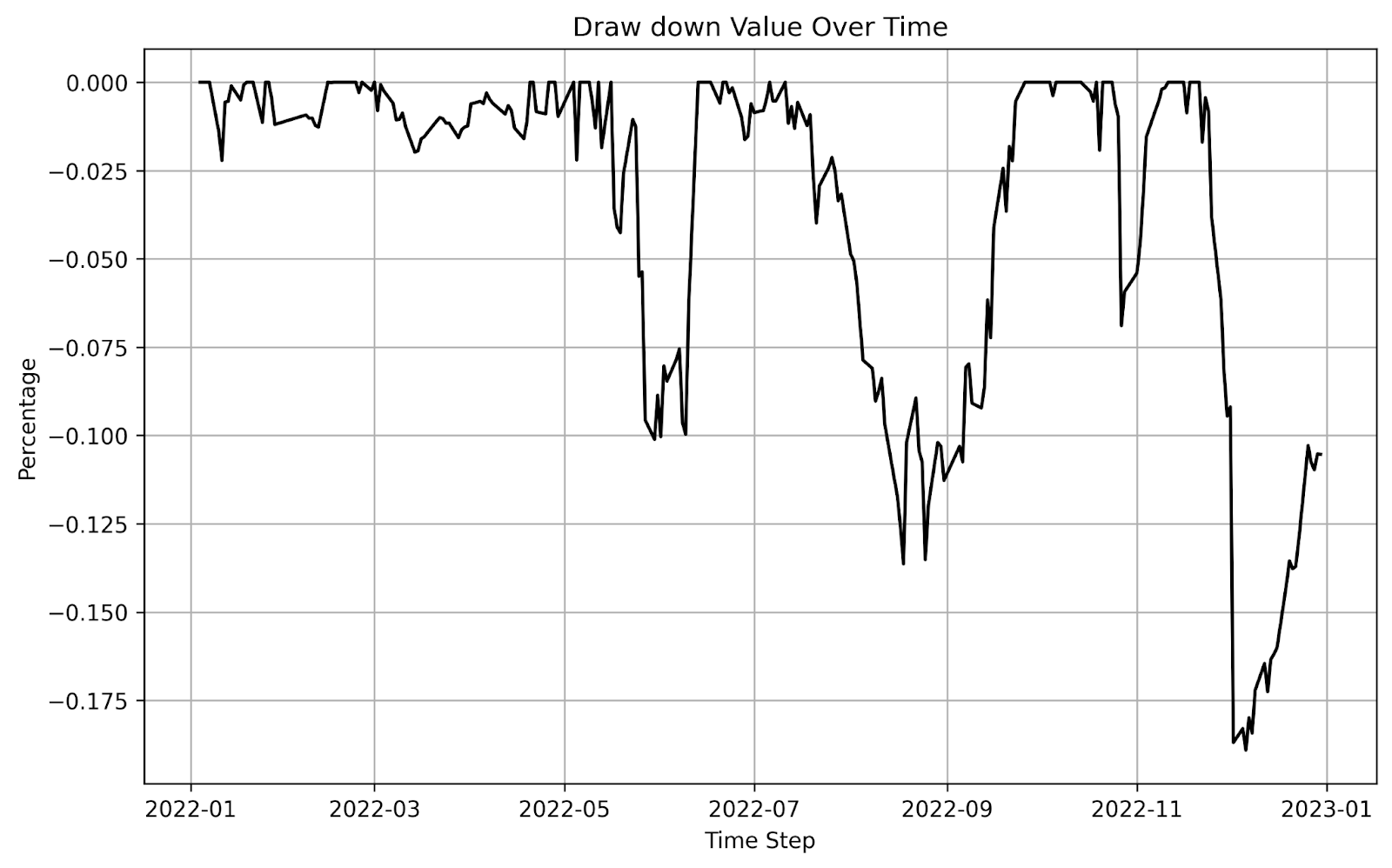}
    \caption{Market Maker Drawdown Chart In-sample}
    \label{fig:dd-in-mm}
\end{figure}
\begin{figure}[ht]
    \centering
    \includegraphics[width=\columnwidth]{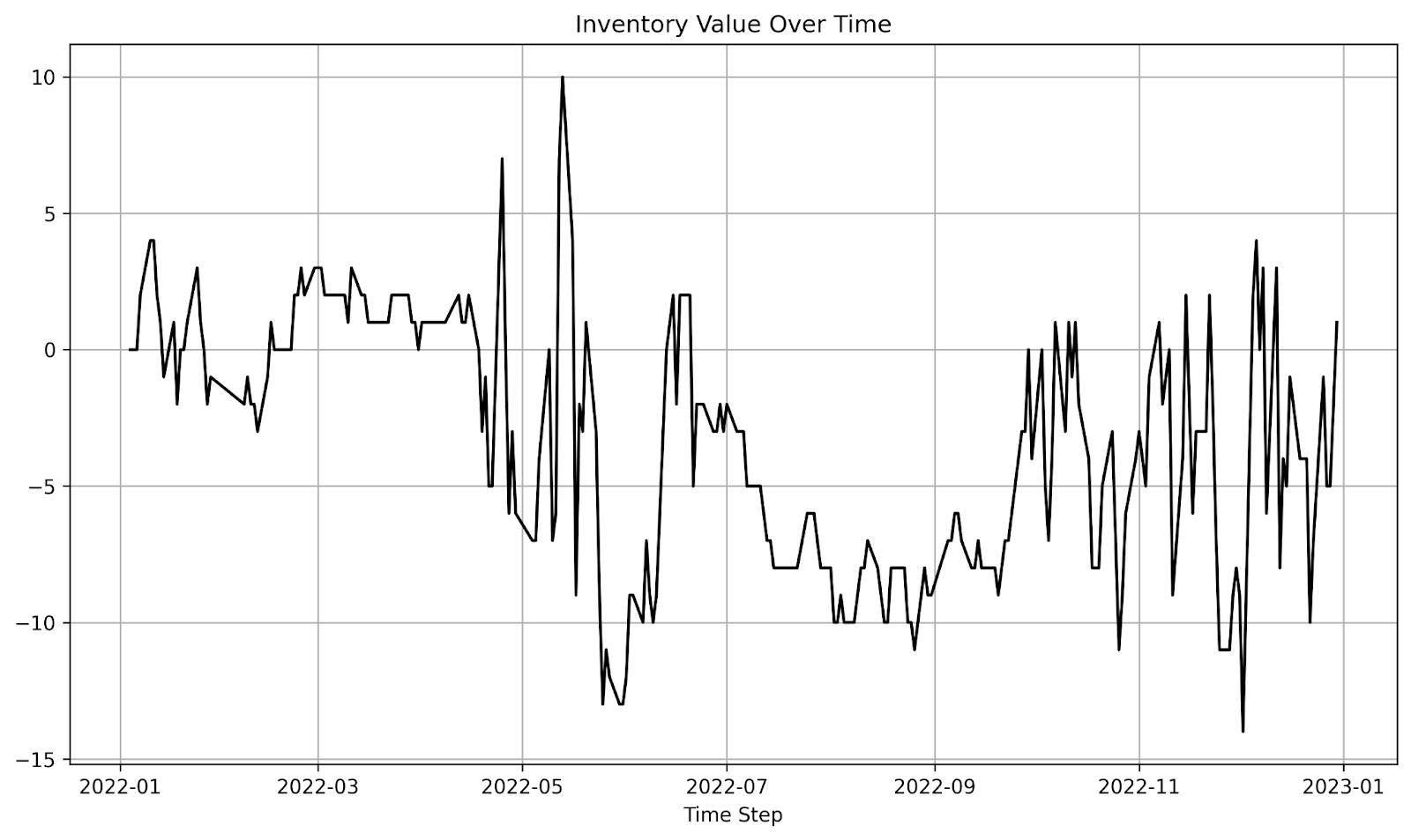}
    \caption{Market Maker Daily Inventory Chart In-sample}
    \label{fig:di-in-mm}
\end{figure}

\subsection{Optimization}
The step size parameter controls the trade-off between fill probability and profit per trade.

A randomized optimization routine was run using seed 2025, with the objective function set to maximize the Sharpe Ratio during the in-sample period. This ensured the resulting configuration balanced profitability and risk across varying inventory exposures.

\begin{itemize}
    \item Step Size: $2.940955612440289$
\end{itemize}

This value balanced stability, fill frequency, and fee tolerance during the in-sample phase.

\subsection{Out-of-Sample Backtesting}
\begin{itemize}
    \item Period: Jan 2, 2024 – Apr 29, 2025
    \item Config: Step size from prior optimization retained
    \item Market Regime: Slightly higher volatility, mild directional bias
\end{itemize}

\subsubsection{Performance Metrics}
In the out-of-sample phase, the strategy struggled to maintain profitability. Both the Sharpe and Sortino ratios turned negative, reflecting weaker risk-adjusted returns and more persistent drawdowns. Nevertheless, the system’s internal mechanics, such as inventory controls and execution cadence, remained stable, and the maximum drawdown of $\sim 21\%$ shows that risk was still bounded.
\begin{table}[!ht]
\centering
\begin{tabular}{ll}
\textbf{Metric}        & \textbf{Value} \\
Sharpe Ratio           & -0.0536        \\
Sortino Ratio          & -0.0673        \\
Maximum Drawdown (MDD) & -21.37\%
\end{tabular}
\caption{In-sample Backtesting Performance}
\label{tab:market-maker-out-of-sample}
\end{table}

\subsubsection{Charts}
The NAV curve shows a mostly sideways and slightly deteriorating trajectory. The drawdown curve reflects choppier equity performance, while the inventory plot remains stable, suggesting that the quoting logic still functioned as designed even under less favorable market conditions.
\begin{figure}[ht]
    \centering
    \includegraphics[width=\columnwidth]{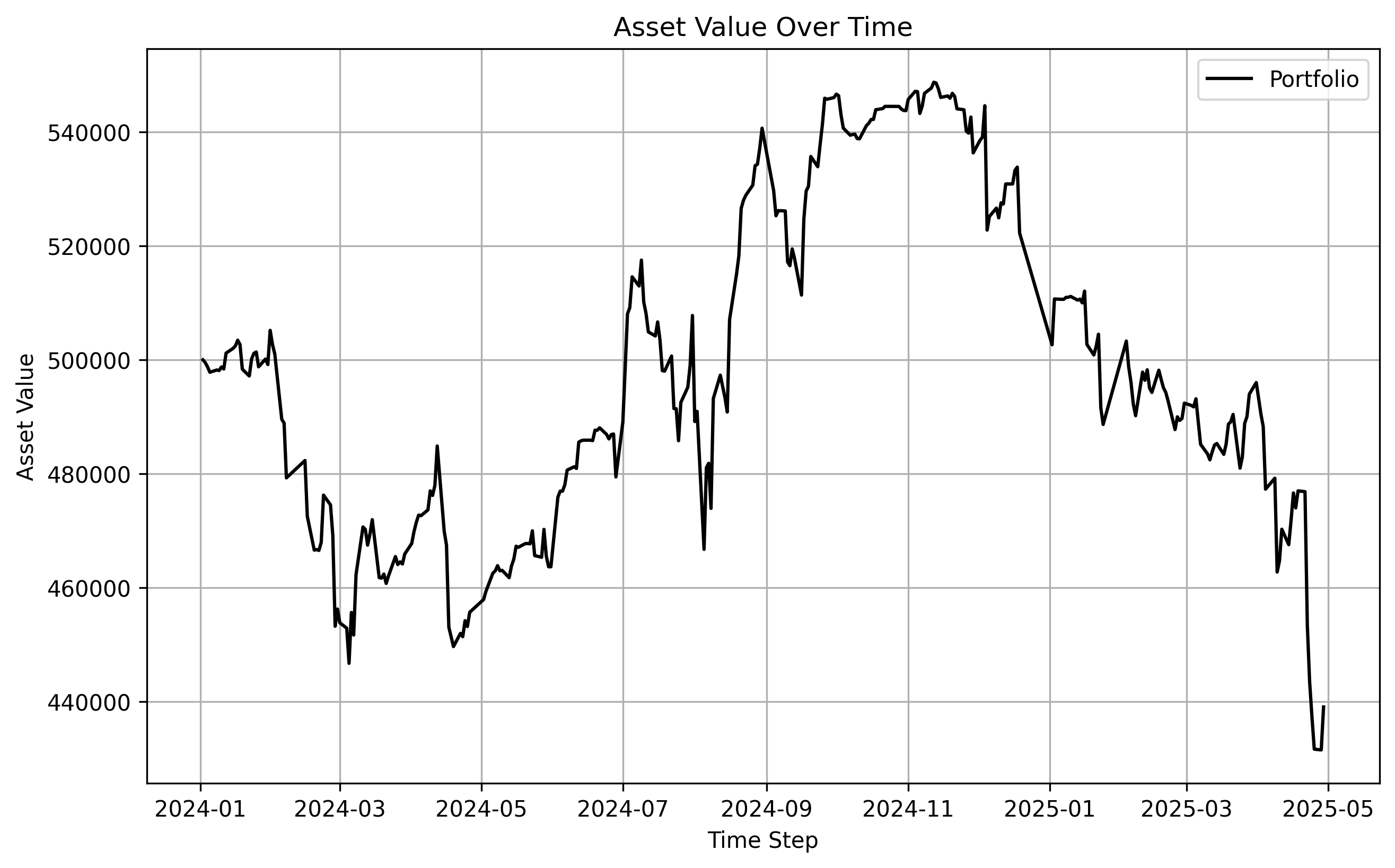}
    \caption{Market Maker NAV Chart Out-of-Sample}
    \label{fig:nav-out-mm}
\end{figure}
\begin{figure}[ht]
    \centering
    \includegraphics[width=\columnwidth]{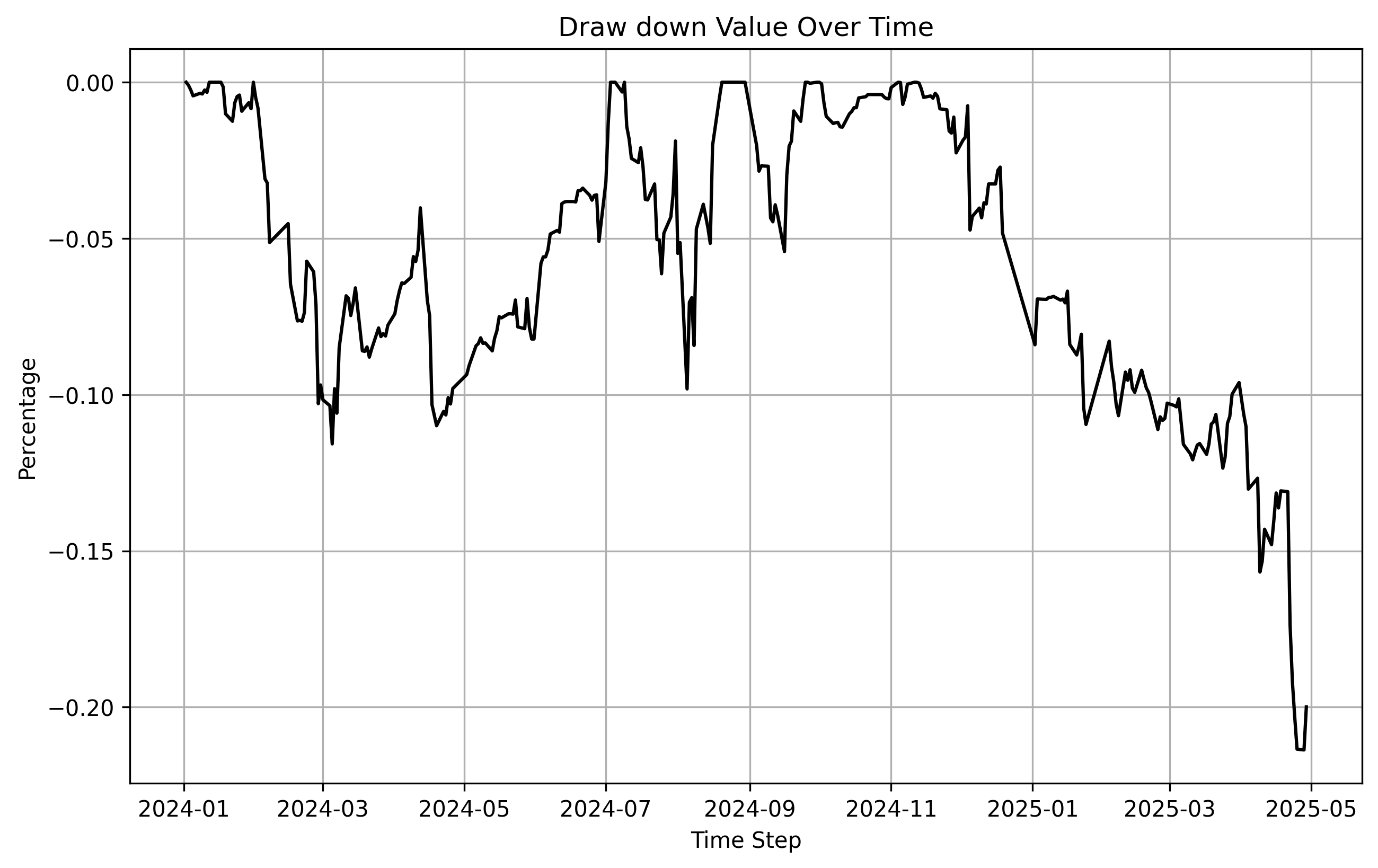}
    \caption{Market Maker Drawdown Chart Out-of-Sample}
    \label{fig:dd-out-mm}
\end{figure}
\begin{figure}[ht]
    \centering
    \includegraphics[width=\columnwidth]{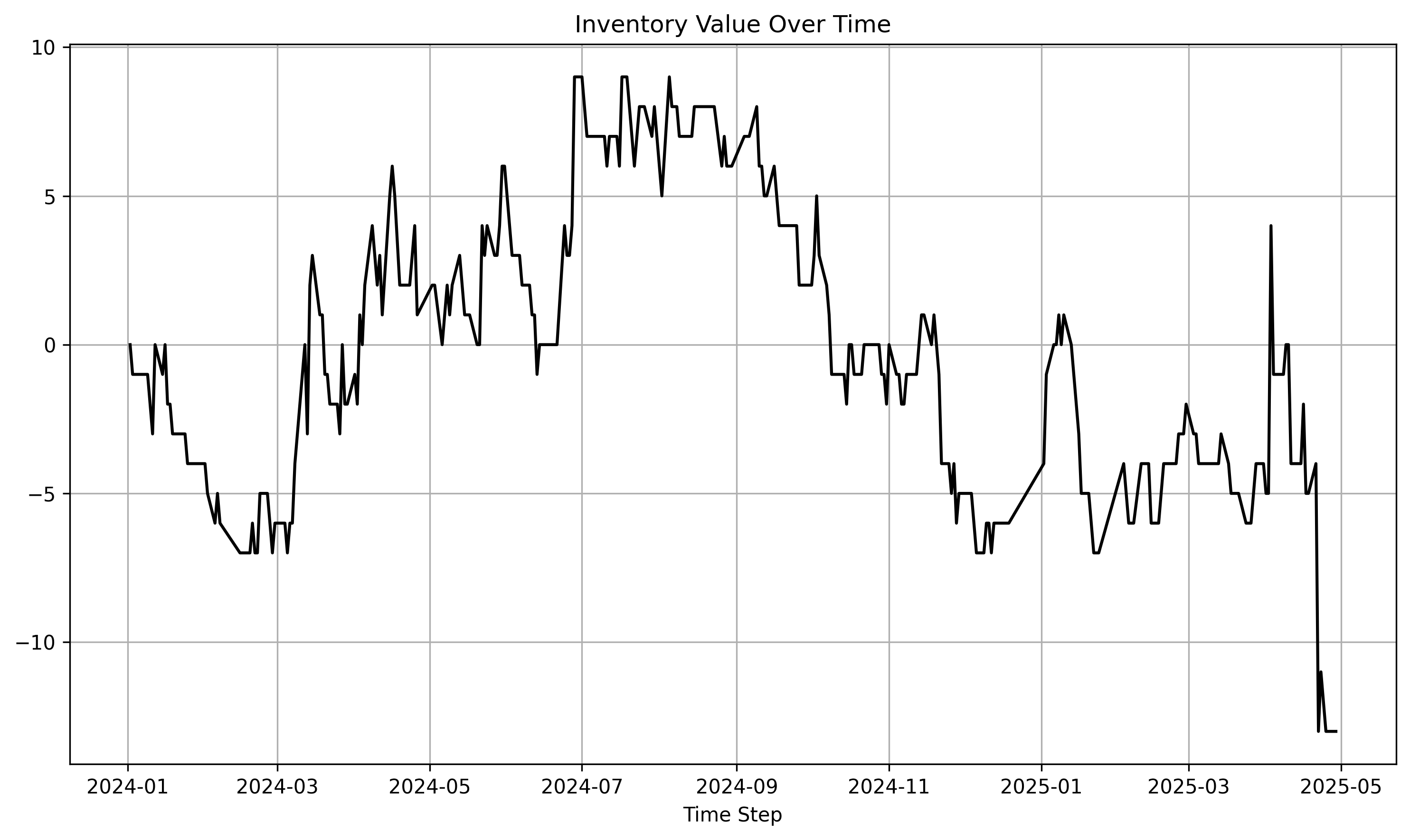}
    \caption{Market Maker Daily Inventory Chart Out-of-Sample}
    \label{fig:di-out-mm}
\end{figure}

\subsection{Paper Trading}
No live market deployment has been conducted yet. This strategy is currently validated only through historical simulation.

\subsection{Conclusion}
PROTO: Market Maker illustrates that reproducibility can extend into the domain of fast-moving execution strategies. A PLUTUS-compliant implementation can make trading logic transparent, measurable, and extensible even when faced with tick-level complexity and sensitive parameters.

Together with PROTO: Smart Beta, this strategy rounds out the initial PROTO series, a growing suite of reference implementations designed to serve as launchpads for deeper exploration and collaboration.

\section{Contributing to PLUTUS: Building Together}
PLUTUS is not just a framework. It is a movement. Like any open-source ecosystem, its long-term success hinges not on a single company but on the collective effort of contributors worldwide.

This paper presents PLUTUS as a structured, reproducible, and transparent approach to algorithmic trading research. But its full potential will only be realized when builders of all backgrounds engage with it by using, extending, testing, and evolving the ecosystem.

Whether you are a quant developer, a student writing your first strategy, or a finance professional looking to scale your ideas, PLUTUS has a role for you.

\subsection{Why Contributions Matter}
The problems PLUTUS addresses, fragmentation, irreproducibility, and opacity, are not just technical. They are systemic. Fixing them requires more than software. It requires a culture change.

When you contribute to PLUTUS, you help:
\begin{itemize}
    \item Raise the standard for trading research quality
    \item Make learning and collaboration easier for others
    Create momentum around open best practices
    \item Expand the diversity of ideas and strategies available in the field
\end{itemize}

In short, you don't just add code. You help reshape the ecosystem.

\subsection{Ways to Contribute}
\subsubsection{Develop PLUTUS-Compliant Projects}
Apply the reproducibility standard to your algorithmic strategies. Use the README.md structure, follow the 9-step methodology, and report your results. This is the easiest and most impactful way to participate, by showing what good looks like.

\subsubsection{Contribute to the PLUTUS Framework}
Help improve the shared infrastructure. You can:
\begin{itemize}
    \item Fix bugs or optimize components
    \item Add new strategy templates, data loaders, or evaluators
    \item Improve documentation and developer onboarding
    \item Integrate external tools, APIs, or execution engines
\end{itemize}
Think of this as DevOps for quantitative finance.

\subsubsection{Refine the PLUTUS Standard}
The reproducibility rubric isn't static. You can:
\begin{itemize}
    \item Propose improvements to the documentation structure
    \item Suggest better evaluation metrics or compliance scoring
    \item Stress-test edge cases through unconventional strategies
\end{itemize}
The standard should evolve alongside the community.

\subsubsection{Improve Sample Projects}
The PROTO series is where many newcomers start. You can:
\begin{itemize}
    \item Add test cases
    \item Expand across markets or asset classes
    \item Fix broken steps or outdated dependencies
    \item Translate documentation for broader access
\end{itemize}

These are low-barrier, high-impact contributions.

\subsubsection{Join the Conversation}
Collaboration doesn't just happen in code. Join forums, review pull requests, host study groups, write explainers, or mentor others entering the field.

PLUTUS is a tool, but it's also a conversation. And conversations need participants.

\subsection{From Users to Stewards}
Every open project begins with a small group of maintainers. But the best ones outgrow them. Our goal with PLUTUS is to create a system that scales beyond us.

By contributing today, you're helping build:
\begin{itemize}
    \item A globally accessible body of trading knowledge
    \item A shared protocol for evaluating and exchanging ideas
    \item A foundation for more credible, inclusive, and accelerated innovation
\end{itemize}

Whether you publish your first backtest or are architecting multi-strategy portfolios, your voice matters. The future of algorithmic trading doesn't belong behind paywalls or in black boxes. It belongs to the people who built it together.

\section{Looking Ahead}
PLUTUS Open Source began as a response to the silence, the locked doors, the unverifiable results, the isolated notebooks. But it has quickly become something much more: a blueprint for how algorithmic trading can evolve when openness becomes a first principle.

This paper presents PLUTUS as a reproducible, community-driven framework for algorithmic trading research. It introduces a practical standard, a modular development process, and a growing body of reference projects, all designed to encourage transparency, rigor, and collaboration in a domain that has long resisted all three.

But PLUTUS is not finished. It is barely beginning.

The roadmap ahead includes:
\begin{itemize}
    \item Expansion of the PROTO series to cover more strategy classes: arbitrage, grid trading, volatility overlays, multi-asset models, and more
    \item Development of a hosted platform for strategy testing, compliance scoring, and paper trading integration
    \item Once strategies have been developed and rigorously validated using the PLUTUS platform and standard, they can be seamlessly integrated with any brokerage that supports PLUTUS Open Source, allowing practitioners to deploy and realize their discovered alphas with minimal friction
    \item Creation of learning tracks for both technical and financial audiences, helping more people break into the field with confidence
    \item Continued refinement of the standard as community feedback grows
\end{itemize}

Most importantly, PLUTUS will grow through participation.

This project is for you if you have ever struggled to verify a strategy, explain your logic, or find a credible path into the field. If you've ever wanted to teach, share, or build something you're proud of, this community is for you.

Let us build the future of algorithmic trading, not in secret, but in the open.

\bibliographystyle{ieeetr}  
\bibliography{ref}      

\begin{thebibliography}{10}

\bibitem{bailey2014overfitting}
D.~H. Bailey, J.~M. Borwein, M.~López~de Prado, and Q.~J. Zhu, ``The probability of backtest overfitting,'' {\em Journal of Computational Finance}, vol.~20, no.~4, pp.~39--69, 2014.

\bibitem{ioannidis2005false}
J.~P. Ioannidis, ``Why most published research findings are false,'' {\em PLoS Medicine}, vol.~2, no.~8, p.~e124, 2005.

\bibitem{openscience2015psych}
O.~S. Collaboration, ``Estimating the reproducibility of psychological science,'' {\em Science}, vol.~349, no.~6251, p.~aac4716, 2015.

\bibitem{baker2016repro}
M.~Baker, ``1,500 scientists lift the lid on reproducibility,'' {\em Nature}, vol.~533, no.~7604, pp.~452--454, 2016.

\bibitem{abadi2016tensorflow}
M.~Abadi {\em et~al.}, ``Tensorflow: Large-scale machine learning on heterogeneous systems.'' \url{https://www.tensorflow.org}, 2016.

\bibitem{paszke2019pytorch}
A.~Paszke, S.~Gross, F.~Massa, A.~Lerer, J.~Bradbury, G.~Chanan, T.~Killeen, Z.~Lin, N.~Gimelshein, L.~Antiga, {\em et~al.}, ``Pytorch: An imperative style, high-performance deep learning library,'' in {\em Advances in Neural Information Processing Systems (NeurIPS)}, 2019.

\bibitem{pedregosa2011scikit}
F.~Pedregosa, G.~Varoquaux, A.~Gramfort, V.~Michel, B.~Thirion, O.~Grisel, M.~Blondel, P.~Prettenhofer, R.~Weiss, V.~Dubourg, {\em et~al.}, ``Scikit-learn: Machine learning in python,'' {\em Journal of Machine Learning Research}, vol.~12, pp.~2825--2830, 2011.

\bibitem{quantconnect2024}
QuantConnect, ``Quantconnect platform.'' \url{https://www.quantconnect.com}, 2024.

\bibitem{zipline}
Quantopian, ``Zipline backtesting library (archived).'' \url{https://www.zipline.io}.
\newblock Accessed: 2024-05-20.

\bibitem{backtrader2024}
Backtrader, ``Backtrader documentation.'' \url{https://www.backtrader.com}, 2024.

\bibitem{lopez2018advances}
M.~L{\'o}pez~de Prado, {\em Advances in Financial Machine Learning}.
\newblock Wiley, 2018.

\end{thebibliography}

\end{document}